# Managing Support Requests in Open Source Software Project: The Role of Online Forums

Faheem Ahmed [a], Piers Campbell [a], Ahmad Jaffar [a], Luiz Fernando Capretz [b]

[a] College of Information Technology, United Arab Emirates University

P. O. Box 17551, Al Ain, United Arab Emirates

[b] Department of Electrical & Computer Engineering, University of Western Ontario,
London, Ontario, Canada N6A 5B9

f.ahmed@uaeu.ac.ae, p.campbell@uaeu.ac.ae, ajaffar@uaeu.ac.ae, lcapretz@eng.uwo.ca

**Abstract**

*The use of free and open source software is gaining momentum due to the ever increasing availability and use of the Internet. Organizations are also now adopting open source software, despite some reservations in particular regarding the provision and availability of support. One of the greatest concerns about free and open source software is the availability of post release support and the handling of for support. A common belief is that there is no appropriate support available for this class of software, while an alternative argument is that due to the active involvement of Internet users in online forums, there is in fact a large resource available that communicates and manages the management of support requests. The research model of this empirical investigation establishes and studies the relationship between open source software support requests and online public forums. The results of this empirical study provide evidence about the realities of support that is present in open source software projects. We used a dataset consisting of 616 open source software projects covering a broad range of categories in this investigation. The results show that online forums play a significant role in managing support requests in open source software, thus becoming a major source of assistance in maintenance of the open source projects.*

## 1. Introduction

The term Open-Source Software (OSS) is used to represent free software which gives the user unrestricted access to its source code [1]. Open source is a software development methodology that makes source code available to a large community who participate in the development by following flexible processes and communicating via the Internet [2] .In the recent past many large software development companies are committing themselves to open source projects which has in turn given momentum to this initiative. The favorable acceptance of OSS products by business and the direct involvement of major IT vendors in OSS development have transformed OSS from a fringe activity to the mainstream, [3]. As an economically viable alternative, open source software (OSS) has proven to have reduced maintenance costs, which benefits all stakeholders rather than the traditional model which was dominated by profit making vendors. In achieving better software quality with potentially limitless functionality, there exists no better way to develop software and it is essential that a non-biased mass community collaborative spirit is allowed to flourish.

Active open source projects usually have a well-defined community with common interests which is involved in either continuously evolving the product (or related products) or in using its results [4]. OSS is developed by loosely organized communities of participants located around the world and working over the Internet. Remarkably, most participants contribute without being employed, paid, or recruited by the organization [5]. The use of the Internet further accelerates the popularity and use of free and open source software at an unprecedented rate of growth [5].

The process of software maintenance in OSS is different from the traditional method of software development. In a traditional setting it is generally required to have a software maintenance team present in the organization to provide post operational support. The support varies from operational aspects to change management. If the support requested is change management then the proposed changes need to be either approved or rejected and are later assigned to a defined set of individuals to implement. The decision of approval or rejection is heavily based on the impact of the changes on the overall software application. This scenario is very different to most OSS cases, where it is not mandatory to get the change request approved from any authority. Anyone can propose a change and implement the change by themselves or a request can be floated among participants in the community. In some cases there are moderators present in the OSS project communities who generally supervise the activities of the OSS projects, and which require the change request to be evaluated before approving for implementation. The process of operational support is also similar, anyone can provide operational support against a request and the online forums generally become the point of contact for those seeking assistance. Although a recent trend is for companies that benefit from OSS to have some employees work on OSS, surveys have shown that the majority of such participants are volunteers [6]. The Floss Survey [7] identified many other reasons why developers were involved in OSS development, including becoming part of the open source community, promoting the open source mode of development, supporting the idea of "free" as an alternative to proprietary software, gaining a reputation, and having fun.

It is very difficult to define any particular software development model that an OSS project follows because the activities are heavily dependent on the participants in the project. OSS community participants contribute software content (programs, artifacts, execution scripts, code reviews, comments, and so on) to web sites in each community and communicate their content updates via online discussion forums, threaded email messages, and newsgroup postings [8]. Software support covers broader aspects and provides assistance in fixing defects (identified by the developer or user), new updates, installation issues, and performance issues.

In some OSS projects there are moderators who manage the process of support to a certain extent but in most cases the project is administered by the community of users. This creates a set of volunteer activities which are not controlled and support becomes uncertain. These support issues include; support requests not having been handled by anyone; an identified defect that has not been fixed; and performance issues that have not been handled. Therefore, OSS projects which do not actively provide support to the users have higher tendency of being out of the scope from users perspectives.

## 1.1 Research Motivations & Related Work

The life cycle of software development illustrates that when the software is delivered to the users, any kind of support activity is accommodated under the umbrella of software maintenance. In reality the maintenance of software has a much broader scope and covers many aspects such as perfective, corrective and enhancement. One of the major concerns in OSS projects is the early delivery to the user and that the lifecycle of the software development process is not consistently followed. Vixie [9] finds that in OSS projects the software life cycle activities such as requirement definition, system level and detailed design, unit and system testing and support are not carried out in a manner similar to traditional software engineering. This raises issues concerning dynamic requirements definition and further increases the importance of maintenance activities. A given change in requirements of any software application is a common occurrence in most software development projects and is generally accommodated in perfective and enhancement maintenance.

## 2. Research Model & Hypotheses of the Study

The world has witnessed a rapid growth of OSS development projects due to the increased popularity and use of Internet. This has formed a diverse community of software developers across the globe, that share directly or indirectly knowledge and communicate using online forums. Their needs and interests are also diverse. An indirect measure of success or failure of an OSS project can be considered by the activeness of the associated online forums. An online forum for an OSS project which has a continuous increase in messages shows the interest of people and helps in managing new features. The main objective of the research model used in this study

(shown in Figure 1) is to analyze the association between OSS support and online public forums associated with the OSS project. The main objective of this study is to investigate the answer to the following research question:

**RQ:** Do online public forums help in managing the support activities in OSS projects?

In order to empirically investigate the research question we hypothesize the following:

**H0:** Online public forums help in managing support activities in OSS project.

**H1a:** The Open support requests present in OSS are positively related with mailing list in online forums.

**H1b:** The open support requests in OSS are positively related with the number of messages in online forums.

**H2a:** The close support requests present in OSS are positively related with mailing list in online forums.

**H2b:** The close support requests in OSS are positively related with the number of messages in online forums.

It is important to mention here that we are using the term "open support requests" as any activity such as change in requirements to add more functionality, installation help, bug fixing support etc. and has not been accommodated yet, whereas "close support requests" refers to a one such activity which was implemented.

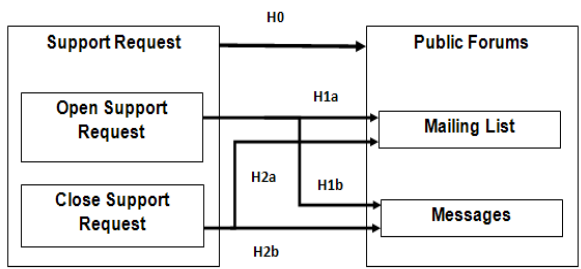

**Figure 1: Research Model of the Study**

## 2.1 Data Collection & Experimental Setup

We collected the data of 1880 open source software projects from www.sourceforge.net, a popular data repository of open source software projects on the internet. The dataset covers various categories of open source software projects such as communication, database, desktop, education, format & protocols, games & entertainment, scientific & engineering, security, software development, system and text editor. The first filtration activity removes the data of all projects which have either a total of new feature requests equal to 0 or that have no online forums. The dataset is therefore reduced from 1880 projects to 650. Subsequently outliers selected on the basis of total new feature request and a number of online forums were removed and this further reduces the dataset to 616 open source projects. Figure 2 illustrates the number of total bugs present in various open source software projects of the initial dataset of this study. It also highlights the outliers which are removed and the updated distribution of total bugs is shown in Figure 3. Figure 4 & 5 illustrates the distribution of total online forums in various open source software projects before and after removing the outliers. In the dataset of this study we used communication (183), database (76), desktop (48), education (21), format & protocols (15), games & entertainment (60), scientific & engineering (41), security (47), software development (28), system (53) and text editor (44) projects. Figure 6 illustrates the distribution of the dataset in various categories.

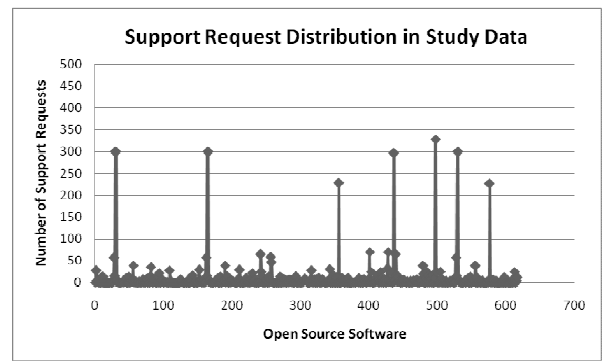

Figure 2: Support request distribution of study dataset

The highest occurrence of open feature requests was found in the category of communication, games/entertainment and scientific and engineering (276). The minimum number of (50) open feature requests were observed in the category of software development. The maximum for feature requests which have been implemented was observed in the categories of database and desktop environment (432 each). The category of system software also shared the minimum number of (100) feature requests which have been implemented. Communication, games/entertainment and scientific and engineering had the observed maximum number (655 each) of total feature requests. The category of software development project had minimum number of (146) known total feature requests. The category of "communication" has the maximum number of online forums (13) in one project. The highest number of messages (5611) was found in the category of communication projects. The lowest number of (2938) messages was observed in the category of database projects. The maximum number of mailing lists of (7) each was observed in the categories of communication, education, games & entertainment and scientific & engineering projects. Desktop environment, security and text editors shared the minimum number of (5) mailing lists in a project.

To analyze the research model and check the significance of hypotheses H1a, H1b, H2a and H2b, we used various statistical analysis techniques. Initially we divided the data analysis activity into three phases. Phase-I dealt with normal distribution tests and parametric statistical analysis. Phase-II dealt with non-parametric statistical analysis. In order to increase the external validity of the study, we used both statistical approaches of parametric and non-parametric methods. We tested for the normal distribution of all six factors of total, open, close bugs as well as number of online forums, mailing list and messages using mean, standard deviation, kurtosis and skewness techniques, and found the values for all these tests to be within the acceptable range for the normal distribution with some minor exceptions. We conducted tests for hypotheses H1a, H1b, H2a and H2b using parametric statistics, such as the Pearson correlation coefficient and one tailed t–test in Phase-I. In Phase-II of non-parametric statistics, we conducted tests for hypotheses using the Spearman correlation coefficient. Phase-III dealt with testing the hypotheses of the research model of this study using the Partial Least Square (PLS) technique. The PLS technique helps when complexity, non-normal distribution, low theoretical information, and small sample size are issues [10] [11]. In the PLS testing of hypotheses we keep one factor as independent and other as dependent variable. We used the PLS technique to increase the reliability of the results. The statistical calculations were performed using Minitab® 14 software.

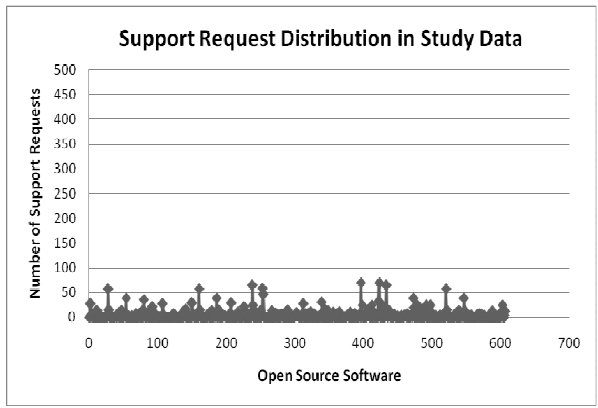

Figure 3: Support request distribution after removing outliers

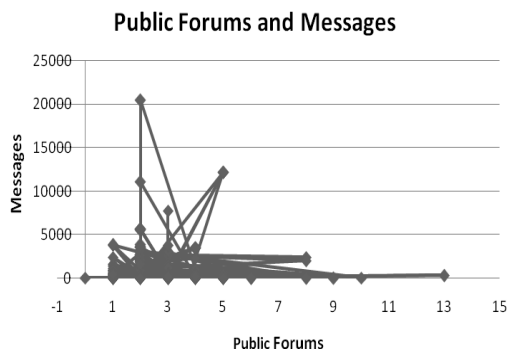

Figure 4: Public forums data distribution

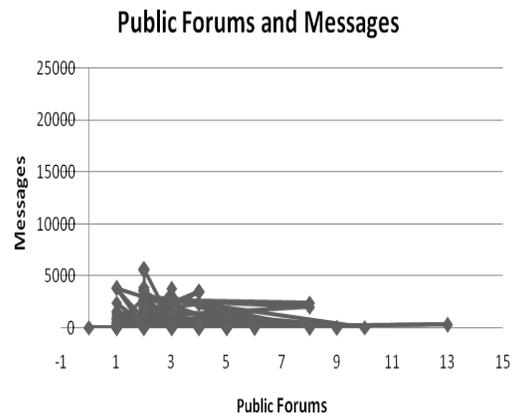

Figure 5: Public forums data distribution after removing outliers

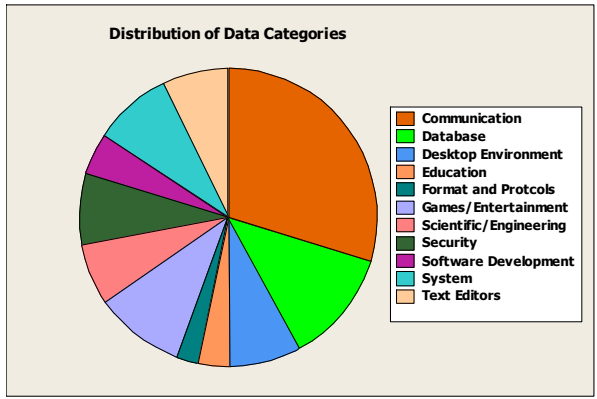

Figure 6: Distribution of project categories in the dataset

## 3. Data Analysis & Results

We examined the Pearson correlation coefficient and t-test between variables involved in the hypotheses H1a, H1b, H2a and H2b. The Pearson correlation coefficient between open support requests and number of mailing lists in the public forums was positive (0.08) at $P < 0.005$, and thus provided a justification to accept the H1a hypothesis. The hypothesis H1b was accepted based on the Pearson correlation coefficient (0.25) at $P < 0.01$, between open support requests and number of messages on the online forum. The correlation coefficient of 0.20 at $P < 0.01$ was observed between the close support requests and number of mailing lists in the online forum. The positive correlation coefficient of 0.23 at $P < 0.01$ meant that H2b was accepted. Hence, it was observed and is reported here that hypotheses H1a, H1b, H2a, and H2b, were found statistically significant and were accepted.

In Phase-II we conducted non-parametric statistical technique using Spearman correlation coefficient to test the hypotheses H1a, H1b, H2a and H2b. Hypothesis H1a was statistically significant at $P < 0.001$ with Spearman correlation coefficient of 0.32. A positive association was observed between open support requests and number of messages (H1b) on the online forum (Spearman: 0.38 at $P < 0.01$). H2a, which deals with between the close support requests and number of mailing lists in the online forum, was accepted (Spearman: 0.29 at $P < 0.01$. The Spearman correlation of (0.45 at $P < 0.01$) was observed for H2b. Hence, it was observed and is reported here that hypotheses H1a, H1b, H2a, and H2b, were found statistically significant and were accepted.

Table 1: PLS Regression Analysis

|  | Open Support Requests | Close Support Requests |
|---|---|---|
| Mailing List | (H1a) Coefficient: 1.93 $R^2$: 0.80 F-ratio: 4.59 P-value: $< 0.03$ | (H2a) Coefficient: 4.01 $R^2$: 0.03 F-ratio: 2.07 P-value: $< 0.001$ |
| Messages | (H1b) Coefficient: 1.01 $R^2$: 0.06 F-ratio: 43.08 P-value: $< 0.001$ | (H2b) Coefficient: 1.02 $R^2$: 0.20 F-ratio: 154.2 P-value: $< 0.001$ |

In Phase-III of hypotheses testing, we used the PLS technique to overcome some of the associated limitations and to cross validate with the results observed using the approaches of Phase-I and Phase-II. We tested the hypothesized relationships, i.e. H1a, H1b, H2a and H2b, by examining their direction and significance. The hypothesis involves two variables therefore in PLS we placed one variable as the response variable and other as the predicate. Table-I reports the results of the structural tests of the hypotheses. It contains observed values of path coefficient, $R^2$ and F-ratio. The path coefficient open support requests (H1a) was found to be 1.93, $R^2$: 0.80 and F-ratio (4.59) was significant at $P < 0.03$. Open support requests (H1b) had positive path coefficient of 1.01 with $R^2$: 0.06 and at $P < 0.001$ F-ratio was 43.08 with number of messages. Close support requests with mailing list (H2a) (Path coefficient: 4.01, $R^2$: 0.03, F-ratio: 2.07 at $P < 0.001$) had the same direction as proposed. Close support requests and number of messages (H2b) (Path coefficient: 1.02, $R^2$: 0.20, F-ratio: 154.2 at $P < 0.001$) also had the same direction as proposed in H2b. All in all, the hypotheses H1a, H1b, H2a and H2b, showed significant at $P < 0.001$ with a positive path coefficient and were in the same direction as proposed.

## 4. Discussion of Empirical Evidence

Open source software is increasing in popularity due to the volume of involvement in its management and use. One of the major reasons behind this advancement is the economic aspects and the other notable reason is the use of the Internet, which has virtually scaled down the world into a knowledge village. The role of support is significantly important in the whole development cycle of the OSS. Notably, the success and failure of an OSS is dependent on the perception and actual availability of support. Typically support is provided through the exchange of messages in the online forum.  These messages are of different types and ranges from support requests, bug identification and fixing, new feature request etc. It is clear from analysis presented that the number of open support requests and the number of online messages are positively related. The OSS community is renowned for its close interaction of professional and amateur software developers and the development characteristics of OSS projects ensure that reuse is a central pillar in project development. As such the user community will request support in the hope that other developers have already encountered the same issues and can provide solutions or recommendations regarding the product.  The development style of Open Source Software is very much user driven and the focus of users in communities such as Sourceforge.net is the continual enhancement of software, particularly when the application in question provides an open alternative to commercially available applications. As a result the interest in such applications is high and the support base drives forward the development of new features which are subsequently included in the successive releases. The main interaction for such requests is via the forum(s) related to the software, however, in order to interact with the forum it is essential that users are registered in the mailing list associated with the software application. As a result and as identified by our analysis, the greater the number of registered mailing list users the higher the number of open support requests. The Open Source Software community is highly active, as clearly shown by the number of projects included in this study and we have already identified that support is correlated to the interaction with mailing list and forums. However our analysis also identifies a positive correlation between both the number of support requests and size of mailing list and the number of closed support requests relating to each application. Again we propose that this correlation is due to the spirit of reuse and collegiality engendered through the open source community. As support is requested via forum messages (and mailing lists) by registered participants in the community, they are quickly serviced by other members, resulting in the closure of the support request. It is this ongoing cycle of requesting, solving (which may include developing and implementing) which makes open source software such an appealing area of software development.

## 5. Conclusion

There are clear indications that free and open source software is gaining in its share of the software market as the quality of the projects grow and users move away from commercial software. Indeed even organizations despite some concerns about quality have been using this type of software for variety of purposes. The objective of this study was to analyze empirically the association between managing support requests OSS projects and the online public forums associated with a given OSS project. We observed that online forums are the corner stone of managing support requests in OSS. The management of such requests from inception, through investigation to closure is communicated via online forums. This study further helps in understanding the significant role of online forums in OSS development. We are currently working on a prediction model to predict the new feature requests in an OSS project based on the active involvement of the online community associated with the projects.